\documentclass[twocolumn,aps,pra,floatfix,showpacs,noshowkeys,epsfig,graphics]{revtex4-1}
\usepackage{graphicx}
\usepackage{amsmath}
\usepackage{amsfonts}
\usepackage{amssymb}
\usepackage{amsthm}

\newcommand{\newc}{\newcommand}
\newc{\beq}    {\begin{equation}}
\newc{\eeq}    {\end{equation}}

\newc{\beqa}    {\begin{eqnarray}}
\newc{\eeqa}    {\end{eqnarray}}
\newc{\bs}    {\section}
\newc{\no}    {\\ \nonumber}

\newc{\st}    {\stackrel}

\begin{document}
\title{ Quantum entanglement from the holographic principle }
\author{Jae-Weon Lee}\email{scikid@jwu.ac.kr}
\affiliation{ Department of energy resources development,
Jungwon
 University,  5 dongburi, Goesan-eup, Goesan-gun Chungbuk Korea
367-805}

\date{\today}

\begin{abstract}
It is  suggested
that  quantum entanglement  emerges from the holographic principle stating that
all of  the information of a region (bulk bits)  can be described by the bits
  on its boundary surface. There are  redundancy  and   information loss in the bulk bits
  that  lead to the nonlocal correlation among the bulk bits.
  Quantum field theory overestimates the independent degrees of freedom in the bulk.
  The maximum entanglement in the universe increases as the size of the  cosmic horizon
  and this could be related with the arrow of time and dark energy.
 \end{abstract}

\maketitle


The  nonlocal quantum correlation (quantum entanglement)
 is nowadays widely treated as the valuable physical resource
 exploited in quantum information processing applications such as quantum key distribution and
quantum teleportation~\cite{nielsen}.
However, the  origin of this mysterious phenomenon is  unknown.
If states of a composite system can be  transformed into
the product form $|\psi_1\rangle |\psi_2\rangle$ by a basis transformation, the state are called separable.
A quantum system is entangled, if its state is not separable.
Note that the superposition rule of quantum mechanics does not automatically guarantees the existence of quantum entanglement.
Actually, deciding the separability of a given superposed quantum state is one of the
 most important open problem in quantum information science.

On the other hand, the holographic principle~\cite{hooft-1993,susskind-1995-36}, including the AdS/CFT correspondence~\cite{Maldacena},
asserts a mysterious  connection between the physics in a bulk
and quantum field theory (QFT) on its boundary surface.
It claims
 that all of  the information in a volume  can be described by the
 degrees of freedom (DOF)
  on the boundary
 of the volume  and
the number of
bits $N_B$ (times ln 2) involved in the description of the bulk must
not exceed $A/4$, where $A$ is the area of the boundary~\cite{Bousso:2002ju}.
Recently, there are renewed interests in describing gravity
with thermodynamics and holography~\cite{Padmanabhan:2009vy,Verlinde:2010hp}.

There is an unexpected
 similarity between  entanglement and the holographic principle.
For example, in general, entanglement entropy has an  area law and the holographic principle involves
nonlocality by nature.
 Furthermore, it is possible to study black hole entropy using entanglement
entropy~\cite{ryu:181602,emparan-2006-0606,solodukhin:201601,PhysRevD.34.373,hckim}.
Ryu and  Takayanagi  proposed a holographic derivation of the entanglement entropy using the AdS/CFT correspondence~\cite{Ryu:2006bv}.
Interestingly, a superluminal (i.e., faster than light) communication is impossible even with the quantum nonlocality.
These counterintuitive results imply that
gravity and quantum mechanics somehow cooperate not to violate each other
and there is a deep connection between them.

In this paper, we suggest that quantum entanglement emerges from
 holographic principle. 't Hooft proposed that quantum mechanics
 has a deterministic theory involving local information loss~\cite{hooft-2002}.
 Zeilinger and Brukner ~\cite{zeilinger1999,brukner-1999-83} suggested that
 every  well-designed experiment tests some proposition which may return a yes/no answer, and
 quantum randomness arises from this discreteness of information. 
Inspired by the digital nature of the  holographic principle
we assume that both of the  bulk and the boundary DOF can be treated as   binary variables.
 We also restrict ourselves to pure states for simplicity. Introducing mixed states does not change main conclusions.

Consider a spherical bulk region $\Omega$ in Fig. 1 with radius $R$
and a causal horizon $\partial \Omega$ such as a black hole horizon.
 An inside observer $\Theta_I$ can describe the bulk physics with bulk bits $B_\alpha$, whereas,
according to the holographic principle, an outside observer $\Theta_O$ can fully describe the bulk physics
using only  the  bits $b_i$ on the boundary.
Starting from the holographic principle, it is simple to show the existence of quantum entanglement of the bulk
quantum states by a
proof by contradiction as follows. \\
i) Assume there is no entanglement  at all in the bulk.
ii) Then, all possible bulk states are product states $|B_1\rangle|B_2\rangle\cdots|B_\beta\rangle$ by the definition of the entanglement.
iii) The number of independent bulk bits $N_B$ is volume proportional (i.e., $O(R^3)$), while the number of boundary bits $N_b$ is of $O(R^2)$.
iv) Thus, one can not fully describe the bulk physics using only the boundary bits $b_i$.
v) This is contradictory to the holographic principle; therefore
 there should be some entangled states. 


\begin{figure}[tpbh]
\includegraphics[width=0.4\textwidth]{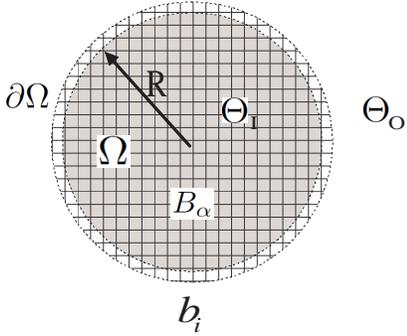}
\caption{Consider a spherical bulk region $\Omega$ with a radius $R$ and a causal horizon $\partial \Omega$.
An inside observer $\Theta_I$
 can describe the bulk physics with bulk bits $B_\alpha$, while, according to the holographic principle, an outside observer $\Theta_O$ can describe the bulk physics
with the  bits $b_i$ on the boundary fewer than $B_\alpha$.
This redundancy could be the origin of the quantum entanglement of the bulk quantum states.}
\end{figure}


Then, exactly how entanglement arises?
According to the holographic principle there is redundancy in the bulk bits $B_\alpha$.
They are not independent of each other.
Simply ignoring some bulk bits could not be a solution, because 
the boundary bits should be able to reproduce arbitrary configuration of the bulk bits,
at least probabilistically. Therefore, only possible way seems to be $n$ to $1$ correspondence between the bulk bits and the boundary bits.
Mathematically, this could mean that there
is a $2^{N_B}$ to $2^{N_b}$  mapping
$f:2^{N_B}\rightarrow 2^{N_b}$.
Since the boundary bits should fully describe the bulk bits (at least probabilistically),
this mapping is a surjective function.

As a toy example, consider a combination of two bulk bits $B_1$ and $B_2$ which is described by
a common boundary bit $b_0$ such that
 both of $(B_1,B_2)=(0,0)$ and $(B_1,B_2)=(1,1)$ correspond to $b_0=0$
and both of $(B_1,B_2)=(1,0)$ and $(B_1,B_2)=(0,1)$ correspond to $b_0=1$. Some information in the bulk bits is lost during the mapping.
(This reminds us of the information loss process considered by 't Hooft
in the quantum determinism proposal~\cite{hooft-2002}.
He  introduced equivalence classes of states
that  evolve into one and the same
state.)
Now, assume that $b_0=1$.
This specific  mapping  can be represented by a matrix relation
\begin{eqnarray}
           \begin{bmatrix}
            0  \\
            1  \\
            1  \\
            0
           \end{bmatrix}_B
           =
             \begin{bmatrix}
            1 & 0  \\
            0 & 1  \\
            0 & 1  \\
            1 & 0
           \end{bmatrix}
           \begin{bmatrix}
            0   \\
            1
           \end{bmatrix}_{b},
\end{eqnarray}
where the vector on the left represents the bulk bits in the basis $(00,01,10,11)$
and the vector on the right represents the boundary bits in the basis $(0,1)$, respectively.
The 4 by 2 matrix represents $f$.
With only $b_0$ value
the outside observer can not distinguish two cases
$(B_1,B_2)=(1,0)$ and $(B_1,B_2)=(0,1)$.
Thus, the statistical probability of $b_0$ estimated by the outside observer should be an addition
of two probabilities,
\beq
P_b=P_B((1,0))+P_B((0,1)),
\eeq
where $P_B((1,0))=1/2=P_B((0,1))$ is the probability that  $(B_1,B_2)=(1,0)$
and $P_b=1$ is the probability that $b_0=1$.
In  the path integral formalism,
for the inside observer this probability could correspond to an entangled quantum state
\beq
\psi=\frac{1}{\sqrt{2}}(|1\rangle|0\rangle+|0\rangle|1\rangle).
\eeq
 In Ref. \cite{myquantum}  it was shown that  quantum mechanics
is not fundamental but emerges from  the information loss at causal horizons,
and that this approach even can lead to the holographic principle~\cite{myholography}.
In the information theoretic formalism, a quantum state in the bulk corresponds to
a statistical probability like $P_B$ estimated by the outside observer who sees the causal horizon. 
This formalism supports the correspondence between $P_b$ and $\psi$. 

We saw that quantum entanglement is unavoidable, once we accept the holographic principle.
The conventional QFT overestimates  DOF in the bulk than are actually present.
Then, how can we reconcile this factor with the great success of the conventional QFT?
Analysis in this work indicates that QFT is valid only for small scales (like particle accelerator scales) compared to the horizon size.
This means that, for example, to study cosmology at the large scale we should not fully trust the result of the conventional
QFT. Dark energy could be a good  example.
It is well known that the zero point energy calculated from the quantum
vacuum fluctuation  is
too large
compared to the observed dark energy.
However, if we invoke the holographic principle and consider only the actual independent DOF of $O(R^2)$,
 the zero point  energy can be comparable to the observed dark energy~\cite{Lee:2010ew}
 and this could resolve the cosmological constant problem.
In other words, there are only $O(R^2)$ independent harmonic oscillators in the bulk QFT.
Our theory predicts that in the bulk there should be always entangled states.
The inside observer can make some of the quantum states separable
but not all of the states {\it at the same time}, because the inside observer cannot remove the redundancy.

Another interesting implication of our theory is that there are $O(R)$ redundancy in the bulk bits
and hence at least $O(R)$ entanglement among the bits. This fact implies that the total entanglement inside
an expanding horizon increases as time goes.
If the causal horizon is the cosmic event horizon  expanding with time, we can say
that this increase of entanglement is related to the arrow of time~\cite{Lee:2010fg}.
Note that this entanglement is different from the entanglement between inside and outside
DOF of the horizon which is usually of $O(R^2)$.

The nonlocality of quantum entanglement is also intimately related to that of the holographic principle.
Since the size of the bulk bits are always  larger than that of the corresponding  boundary bits,
some of the correlated bulk bits  should be spatially further separated
than the boundary bits do.
Thus, even if the boundary bits have the locality, the corresponding bulk bits apparently do not.
However, even in this case, entanglement does not allow superluminal communication,
 because the inside observer cannot choose the specific outcome of her/his measurements.
Neither the outside observer do influence the bulk bits faster than light.
For a fixed outside observer seeing the causal horizons,
due to a large redshift, it takes infinite time for observer's influence
to reach the horizons. Alternatively, if the outside observer free falls to reach the horizon,
 the horizon will disappear and the observer cannot access the boundary bits properly.
 Both of the holography and entanglement are observer dependent phenomena.

In summary,
this paper shows that quantum entanglement is directly related to
the holographic principle.
In a series of paper\cite{Lee:2010xv,myquantum,myholography}, it was shown that quantum mechanics, Einstein gravity and even holographic
principle can be derived by considering phase space information loss at causal horizons.
The analysis in this paper enhance this viewpoint.

\section*{acknowledgments}
This work was supported in part by Basic Science Research Program through the
National Research Foundation of Korea (NRF) funded by the ministry of Education, Science and Technology
(2010-0024761) and the topical research program (2010-T-1) of Asia Pacific Center for Theoretical
Physics.
%

\begin{thebibliography}{10}

\bibitem{nielsen}
M.~A. Nielsen and I.~L. Chuang,
\newblock {\em Quantum Computation and Quantum Information} (Cambridge
  University Press, Cambridge, 2001).

\bibitem{hooft-1993}
G.~'t~Hooft,
\newblock {\em Salam-festschrifft} (World Scientific, Singapore, 1993).

\bibitem{susskind-1995-36}
L.~Susskind,
\newblock J. of Math. Phys. {\bf 36}, 6377 (1995).

\bibitem{Maldacena}
O.~Aharony, S.~S. Gubser, J.~Maldacena, H.~Ooguri, and Y.~Oz,
\newblock Phys. Rep. {\bf 323}, 183 (2000).

\bibitem{Bousso:2002ju}
R.~Bousso,
\newblock Rev. Mod. Phys. {\bf 74}, 825 (2002), hep-th/0203101.

\bibitem{Padmanabhan:2009vy}
T.~Padmanabhan,
\newblock arXiv:0911.5004  (2009).

\bibitem{Verlinde:2010hp}
E.~P. Verlinde,
\newblock arXiv:1001.0785  (2010).

\bibitem{ryu:181602}
S.~Ryu and T.~Takayanagi,
\newblock Phys. Rev. Lett. {\bf 96}, 181602 (2006).

\bibitem{emparan-2006-0606}
R.~Emparan,
\newblock JHEP {\bf 0606}, 012 (2006).

\bibitem{solodukhin:201601}
S.~N. Solodukhin,
\newblock Phys. Rev. Lett. {\bf 97}, 201601 (2006).

\bibitem{PhysRevD.34.373}
L.~Bombelli, R.~K. Koul, J.~Lee, and R.~D. Sorkin,
\newblock Phys. Rev. D {\bf 34}, 373 (1986).

\bibitem{hckim}
M.-H. Lee, H.-C. Kim, and J.~K. Kim,
\newblock Phys. Lett. B {\bf 388}, 487 (1996).

\bibitem{Ryu:2006bv}
S.~Ryu and T.~Takayanagi,
\newblock Phys. Rev. Lett. {\bf 96}, 181602 (2006), hep-th/0603001.

\bibitem{hooft-2002}
G.~'t~Hooft,
\newblock Determinism beneath quantum mechanics, 2002, quant-ph/0212095.

\bibitem{zeilinger1999}
A.~Zeilinger,
\newblock Found. Phys. {\bf 29}, 631 (1999).

\bibitem{brukner-1999-83}
C.~Brukner and A.~Zeilinger,
\newblock Physical Review Letters {\bf 83}, 3354 (1999).

\bibitem{myquantum}
J.-W. Lee,
\newblock Found. Phys. {\bf 41}, 744 (2011), 1005.2739.

\bibitem{myholography}
J.-W. Lee,
\newblock (2011), 1107.3448.

\bibitem{Lee:2010ew}
J.-W. Lee,
\newblock (2010), arXiv:1003.1878.

\bibitem{Lee:2010fg}
J.-W. Lee, H.-C. Kim, and J.~Lee,
\newblock (2010), arXiv:1002.4568.

\bibitem{Lee:2010xv}
J.-W. Lee,
\newblock (2010), 1003.4464.

\end{thebibliography}

\end{document}